# Effect of solvation shell structure on thermopower of liquid redox pairs

Yuchi Chen[1], Qiangqiang Huang[1], Te-Huan Liu[1], Xin Qian[1*] and Ronggui Yang[1,2*]

[1] School of Energy and Power Engineering, Huazhong University of Science and Technology, Wuhan 430074, China.

[2] State Key Laboratory of Coal Combustion, Huazhong University of Science and Technology, Wuhan 430074, China.

*Corresponding emails: xinqian21@hust.edu.cn; ronggui@hust.edu.cn

## Abstract

Recent advancements in thermogalvanic batteries offer a promising route to efficient harvesting of low-grade heat with temperatures below 100 °C. The thermogalvanic temperature coefficient $\alpha$, usually referred to as effective thermopower, is the key parameter determining the power density and efficiency of thermogalvanic batteries. However, the current understanding of improving $\alpha$ of redox pairs remains at the phenomenological level without microscopic insights, and the development of electrolytes with high $\alpha$ largely relies on experimental trial and error. This work applies the free energy perturbation method based on molecular dynamics simulations to predict the thermogalvanic temperature coefficient of the $Fe^{3+}/Fe^{2+}$ redox pair in aqueous and acetone solutions. We showed that $\alpha$ of the $Fe^{3+}/Fe^{2+}$ redox pair can be increased from 1.5$\pm$0.3 mV/K to 4.1$\pm$0.4 mV/K with the increased acetone to water fraction. The predicted $\alpha$ of $Fe^{3+}/Fe^{2+}$ both in pure water and acetone show excellent agreement with experimental values. By monitoring the fluctuation of dipole orientations in the first solvation shell, we discovered a significant change in the variance of solvent dipole orientation between $Fe^{3+}$ and $Fe^{2+}$, which can be a microscopic indicator for large magnitudes of $\alpha$. The effect of acetone weight fraction in the mixed acetone-water solvent on the $\alpha$ of $Fe^{3+}/Fe^{2+}$ is also studied. Acetone molecules are found to intercalate into the first solvation shell of the $Fe^{2+}$ ion at high acetone fractions, while this phenomenon is not observed in the solvation shell of the $Fe^{3+}$ ion. Such solvation shell structure change of $Fe^{2+}$ ions contributes to the enhanced $\alpha$ at high acetone fractions. Our discovery provides atomistic insights into how solvation shell order can be leveraged to develop electrolytes with high thermogalvanic temperature coefficients.

**Keywords**: Heat recovery; thermogalvanic temperature coefficient; solvation shell structure; molecular dynamics; free energy perturbation

**Introduction**

The tremendous amount of low-grade heat, typically at temperatures below 100 °C, from industry and the environment, is a vast renewable energy source but is usually wasted due to the lack of efficient energy harvesting technology[1, 2]. Due to the small temperature differences between low-grade heat sources and the environment[3-5], conventional heat harvesting technologies such as organic Rankine cycles (ORC) and thermoelectric generators (TEG) face the challenges of scalability and cost-effectiveness. In the past few years, the development of redox electrolytes with high effective thermopowers has attracted intensive interest in developing thermo-electrochemical devices such as thermogalvanic cells and thermally regenerative electrochemical cycled batteries using redox-active ions as energy and charge carriers for low-grade heat recovery[6-8]. The effective thermopower of redox electrolytes is quantified by thermogalvanic temperature coefficient $\alpha$, defined as the thermal voltage generated per unit Kelvin of temperature difference[9]:

$$\alpha \coloneqq \frac{\Delta E}{\Delta T} = \frac{\Delta S_{rxn}}{nF} \tag{1}$$

where $E$ is the electrode potential of a redox pair, $T$ is absolute temperature and $\Delta S_{rxn}$ denotes the partial molar entropy change of the reduction reaction $O + ne \rightarrow R$ with $n$ as the electron transfer number and $F$ as the Faraday constant, where $O$ and $R$ denote the oxidized and reduced states of the redox pair, respectively. Compared with the solid-state TEG whose thermopower is usually ~100 μV/K[10, 11], the effective thermopower of typical redox pairs such as $Fe^{3+}/Fe^{2+}$, $Fe(CN)_6^{3-}/Fe(CN)_6^{4-}$ is much larger (~1 mV/K) [9, 12] and has been further improved to a few mV/K in the past few years. For example,

Huang et al.[13] discovered that adding redox-inert supporting electrolytes can effectively improve $\Delta S_{rxn}$ and thereby $\alpha$ due to the structural change in the hydrogen bonding network between water molecules. Yu et al.[14] developed a redox couple with high thermopower by introducing guanidinium ions into $Fe(CN)_6^{3-}/Fe(CN)_6^{4-}$, and the thermogalvanic temperature coefficient is improved to -3.7 mV/K from -1.4 mV/K of the pristine aqueous solution due to thermosensitive crystallization effects. Further addition of urea can result in an optimized thermopower as large as -4.2 mV/K[15]. Introducing nonaqueous solvent molecules is another effective way for improving $\alpha$. Electrolytes based on $Co^{3+}/Co^{2+}$ redox pairs in ionic liquids or organic solvents also showed a $\alpha$ of 2.2 mV/K, higher than typical aqueous transition metal redox pairs[16, 17]. Kim et al.[18] and Liu et al.[19] discovered that adding organic solvents to aqueous $Fe(CN)_6^{3-}/Fe(CN)_6^{4-}$ and $Fe^{3+}/Fe^{2+}$ electrolytes can result in high $\alpha$ values of -2.9 mV/K and 2.5 mV/K, respectively. Recently, Inoue et al.[20] performed a systematical study on the solvent effect on the thermopower of $FeCl_3/FeCl_2$ and discovered that $Fe^{3+}/Fe^{2+}$ can reach 3.6 mV/K in acetone solvents, nearly three times the thermopower of the pristine aqueous solution.

However, the current development of redox electrolytes with high $\alpha$ values largely depends on experimental trial and error due to the lack of microscopic models. Even for the simplest case where both species of the redox pair exist in the solution phase, the analysis of measured temperature coefficients $\alpha$ is often based on the Born solvation model[21, 22]. Although the Born solvation model gives a simple expression that can capture the effects of charge valence, ionic radii, and dielectric constant of solvents on the redox entropy change $\Delta S_{rxn}$, the solvent is simply treated as an effective medium with a dielectric screening effect. As a result, details of microscopic solvation structure are neglected, and the original Born solvation model fails to quantitatively describe the solvent dependence of $\Delta S_{rxn}$, unless solvent reorganization effects are included as correction terms[21, 23]. The Born solvation model also fails to explain the enhanced thermopower by introducing inert supporting electrolytes[13].

This work utilizes the free energy perturbation (FEP) method based on molecular dynamics (MD) simulations to predict the partial entropy change $\Delta S_{rxn}$ and the effective thermopower $\alpha$. Using the $Fe^{3+}/Fe^{2+}$ redox pair as an example, we studied the solvent effect on $\alpha$ and the results showed excellent agreement with experimental values with deviations lower than 25%. We then analyzed the fluctuation of dipole orientations in the solvation shell and discovered that the large difference in the variance of solvent dipole orientations between the oxidized and reduced species can be a microscopic indicator for large magnitudes of $\alpha$ in liquid redox electrolytes. The dependence of acetone fraction in $\alpha$ is also investigated. At high weight fractions of acetone ($\geqslant$ 60%), the acetone molecule is found to get inserted into the first solvation shell of the $Fe^{2+}$ ion, while such a phenomenon does not occur in the solvation shell of the high-valence $Fe^{3+}$ ion. This dramatic difference in the solvation shell structure of the $Fe^{2+}$ ion contributes to the enhanced $\alpha$ at high fractions of acetone. Our discovery provides atomistic insights into how solvation shell order can be tuned for future developments of redox electrolytes with high thermogalvanic temperature coefficients.
5

**Results and discussion**

**Prediction of effective thermopower using FEP**

The key to predicting the thermogalvanic temperature coefficient $\alpha$ is to compute the partial entropy change $\Delta S_{rxn}$ of the redox pair. A variety of mechanisms can induce partial entropy change $\Delta S_{rxn}$, including configurational entropy change $\Delta S^{conf}$, phonon entropy change $\Delta S^{phon}$, electron entropy change $\Delta S^e$, solvation entropy change $\Delta S^{solv}$ and others[24, 25]. The configuration entropy change $\Delta S^{conf}$ is attributed to the atomic configuration change between the oxidized species and the reduced species. The configuration entropy change is usually important in intercalation materials such as $Li_xCoO_2$ [24] and $Li_xFePO_4$ [26]: with the intercalation of Li ions, the host material exhibit atomic structural changes such as redistribution of intercalated Li ions and vacancy sites. The phonon entropy change $\Delta S^{phon}$ accounts for the phonon vibration spectra change of the host electrode material before and after the redox reaction[27], while $\Delta S^e$ is related to the electronic structure change due to the redox reaction. $\Delta S^e$ is usually negligible unless metal-insulator phase transition happens in the electrode material[24]. Finally, $\Delta S^{solv}$ is related to solvation structure change before and after the reaction, as shown in Figure 1a. For simple redox pairs such as $Fe^{3+}/Fe^{2+}$, both $Fe^{3+}$ and $Fe^{2+}$ ions exist in the solution phase, and the electrode does not participate in redox reactions (no electrodeposition or intercalation). In this case, only the solvation entropy change $\Delta S^{solv}$ is significant[21]. Replacing partial entropy with solvation entropy, we obtain a simple expression of the thermogalvanic temperature coefficient for simple redox pairs:

$$\alpha = \frac{\Delta S_R^{solv} - \Delta S_O^{solv}}{nF} = \frac{1}{nF}\left(\frac{d\Delta G_O^{solv}}{dT} - \frac{d\Delta G_R^{solv}}{dT}\right) \tag{2}$$

where $\Delta G_i^{solv}$ denotes the solvation free energy of the oxidized ($i = O$) and the reduced ($i = R$) species, which can be related to the solvation entropy through $\Delta S_i^{solv} = -d\Delta G_i^{solv}/dT$. According to Eq. (2), the thermogalvanic temperature coefficient $\alpha$ can be derived from the temperature-dependence of solvation free energies for each species of the redox pairs.

In this work, we utilize the free energy perturbation (FEP) method[28, 29] based on MD simulations to compute $\Delta G^{solv}$ and $d\Delta G^{solv}/dT$ of both $Fe^{3+}$ and $Fe^{2+}$ ions. The basic idea of the FEP method is to introduce a coupling parameter $\lambda$ that can be perturbatively tuned to construct a thermodynamic path corresponding to the ionic solvation process. As shown in Figure 1b, the potential energy $U$ is then expressed as:

$$U(\lambda) = (1 - \lambda)U_0 + \lambda U_1 \tag{3}$$

where $U_0$ corresponds to the potential energy of the undissolved state in which the ion and the solvent have no interaction, and $U_1$ corresponds to the dissolved state. The parameter $\lambda$ is perturbatively increased from 0 to 1, corresponding to a gradual increase in ion-solvent interactions until the final dissolved state $U_1$ is reached. During the FEP computations, a sequence of coupling parameters $\lambda_j = j/N$ with $j = 0, 1, \ldots, N$ is taken to sample the states on the solvation thermodynamic path, and the solvation free energy is calculated as:

$$\Delta G^{solv}(T) = -k_B T \sum_{j=0}^{N-1} \ln\left\langle \exp\left(-\frac{U(\lambda_{j+1}) - U(\lambda_j)}{k_B T}\right)\right\rangle \tag{4}$$

where $\langle \rangle$ denotes the ensemble average. Parametrization of the interatomic potential used in MD simulations is detailed in Methods.

We first verify the accuracy of FEP calculations by comparing the hydration free energy (the solvation free energy in aqueous systems) of simple cations ($K^+$, $Na^+$) and anions ($Cl^-$, $I^-$) with the literature values, as shown in Table I. All free energy values of simple cations and anions show excellent agreement with the existing literature. For the hydration free energy of $Fe^{3+}$ and $Fe^{2+}$, the calculated values in this work are 10% and 20% higher than those reported by Marcus et al.[30] One possible reason for this small deviation in solvation free energy is that water molecules get polarized under the strong electric field of high-valence ions[31], while in this work the solvent molecules are regarded as rigid. However, computing entropy change $\Delta S_{rxn}$ is not affected by the absolute value of solvation free energy, instead it is the relative free energy change between the oxidized and the reduced species that determines the $\Delta S_{rxn}$. Therefore, we compare the ratio between the free energies of $Fe^{3+}$ and $Fe^{2+}$ predicted by FEP and the experimental values[30], showing reasonable agreement with only 6% deviations.

After computing the solvation free energies of both $Fe^{3+}$ and $Fe^{2+}$ at different temperatures, the thermogalvanic temperature coefficient $\alpha$ can be easily obtained by taking the derivative to temperature. As shown in Figure 1c, the solvation free energies of the $Fe^{3+}$ and $Fe^{2+}$ in water solvent from 280 K and 320 K are calculated, and the solvation entropy of both $Fe^{3+}$ and $Fe^{2+}$ is extracted through least-square linear regression. The solvation free energy of $Fe^{3+}$ and $Fe^{2+}$ ions at different temperatures in acetone are included in Supplementary materials (Figure S2). Note that $d\Delta G_i^{solv}/dT > 0$ indicates a negative solvation entropy change $\Delta S^{solv} < 0$, because the solvent molecules become aligned in the solvation shell due to the constraints of the electric field of the ions, whereas the orientations of the solvent molecules are disordered in the undissolved state. We compute the temperature coefficient $\alpha$ of $Fe^{3+}/Fe^{2+}$ in both pure water and acetone using Eq. (2), and we obtained $\alpha = 1.5 \pm 0.3$ mV/K of aqueous solution and $\alpha = 4.1 \pm 0.4$ mV/K of acetone solution, respectively. These values are within ~25% error compared with electrochemical measurements[9, 20, 32] as shown in Figure 1d. The nearly threefold increase in $\alpha$ by replacing water solvent with acetone indicates that solvent engineering could be an effective route to developing redox electrolytes with high thermopower. In the next section, we analyze the microscopic structure of the solvation shell to provide insights on how to improve $\alpha$ of redox pairs.

**Table I**. Free energy of hydration of some common ions at 298 K (kcal/mol)

|  | $K^+$ | $Na^+$ | $Cl^-$ | $I^-$ | $Fe^{2+}$ | $Fe^{3+}$ | Ratio ($Fe^{3+}/Fe^{2+}$) |
|---|---|---|---|---|---|---|---|
| This work | -69.0 | -87.5 | -104.1 | -74.9 | -392.6 | -852.2 | 2.17 |
| Marcus[30] | -70.5 | -87.2 | -81.3 | -65.7 | -439.8 | -1019.4 | 2.31 |
| Roland[33] | -71.2 | -88.7 | -89.1 | -74.3 | / | / | / |
| Asthagiri[34] | -66.0 | -82.7 | -88.0 | -71.9 | / | / | / |

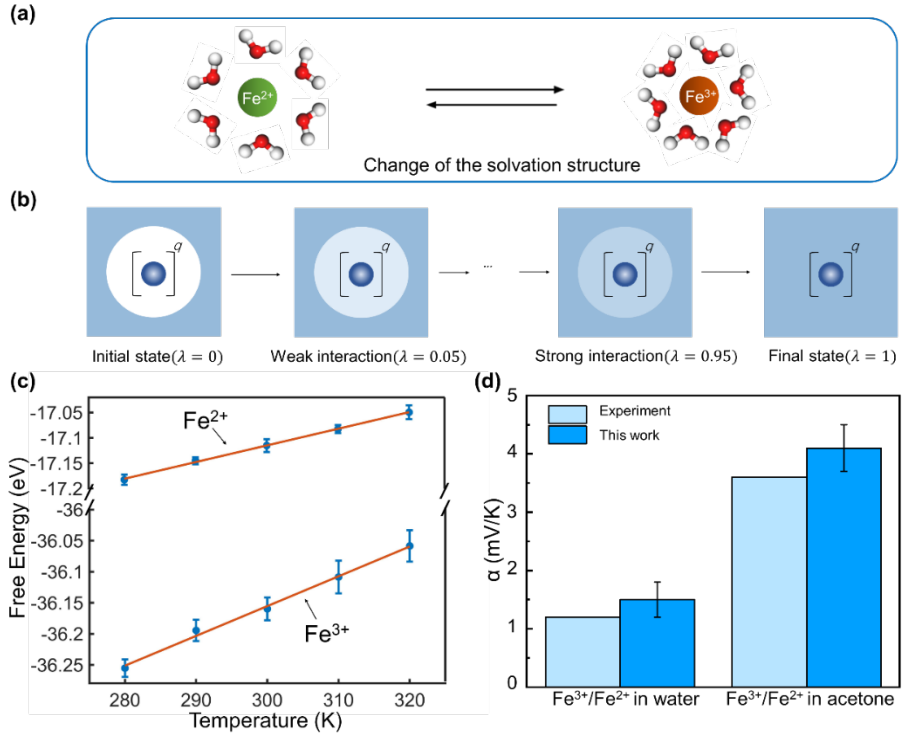

**Figure 1.** (a) Schematic of the solvation structure changes between redox pairs due to valence change. (b) Schematic of the free energy perturbation (FEP) method by gradually changing the coupling parameter $\lambda$. (c) The solvation free energy $\Delta G^{solv}(T)$ of $Fe^{2+}$ and $Fe^{3+}$ in pure water from 280 K to 320 K. The solvation entropy $\Delta S_i^{solv}$ (in eV/K) is obtained from the slope of best linear fit. (d) The predicted temperature coefficients of $Fe^{3+}/Fe^{2+}$ in pure water and pure acetone compared with experimental values[9, 20].

**Analysis of solvation structure**

Understanding the relation between solvation shell structure and $\alpha$ is the key to developing high-$\alpha$ electrolytes[15]. First, we analyze the radial distribution function (RDF) $g(r)$ and the coordination number distribution (CND) $N(r)$ of solvent molecules for the $Fe^{3+}$ and $Fe^{2+}$ ions. As shown in Figure 2a, we simulate a box of 810 water molecules for 10000 ps with a time step of 1 fs to obtain the RDF and CND. Figure 2b clearly shows that the first peak in RDF of the $Fe^{3+}$ ion appears at 1.95 Å, while the first peak locates at 2.05 Å. Such difference in solvation peak is a result of a more tightly bounded solvation shell of $Fe^{3+}$ due to its higher valence compared with $Fe^{2+}$ ion. We can also extract the first solvation shell radius $R_{1st}$ from the position of the first minimum in the ion-oxygen atom RDF[35]. Our MD simulation predicts an $R_{1st}$ of 2.45 Å and 2.55 Å for $Fe^{3+}$ and $Fe^{2+}$ ions, respectively. The 0.1 Å difference in $R_{1st}$ of $Fe^{3+}/Fe^{2+}$ in pure water is in good agreement with the literature[36]. Since there is only one oxygen atom in water and acetone molecules, we can obtain the CND $N(r)$ by directly integrating the Fe-O RDF: $N(r) = \int_0^r g_{\text{Fe-O}}(r')4\pi r'^2 dr'$, where $N(r)$ is the number of solvent molecules bounded with the ion within the cutoff radius $r$. By setting $r = R_{1st}$, it is found that the values of $N(r)$ for both $Fe^{3+}$ and $Fe^{2+}$ is 6.0, indicating that both $Fe^{3+}$ and $Fe^{2+}$ are bonded with 6 water molecules in the first solvation shell, consistent with experimental observations[37].

In acetone solutions (Figure 2c), the $R_{1st}$ of $Fe^{3+}$ and $Fe^{2+}$ is 3.05 Å and 3.35 Å respectively (Figure 2d), ~0.6 Å larger than those in water solutions, which indicates that acetone solvents are more loosely bounded with $Fe^{3+}$ and $Fe^{2+}$ ions. A larger difference of 0.3 Å in $R_{1st}$ is also observed, so that the valence change due to redox reactions of $Fe^{3+}/Fe^{2+}$ pairs would result in a larger solvation structure change, which is consistent with the measured higher $\alpha$ values[19]. The coordination number of acetone in the first solvation shell of $Fe^{3+}$ and $Fe^{2+}$ is 8.4 and 7.8 respectively. Such non-integer coordination number is a result of solvation shell fluctuations, which is commonly observed in MD simulations [38]. The slightly smaller coordination number of $Fe^{2+}$ in acetone than $Fe^{3+}$ is also a sign of stronger fluctuations of the acetone solvation shell than that of the water solvation shell.

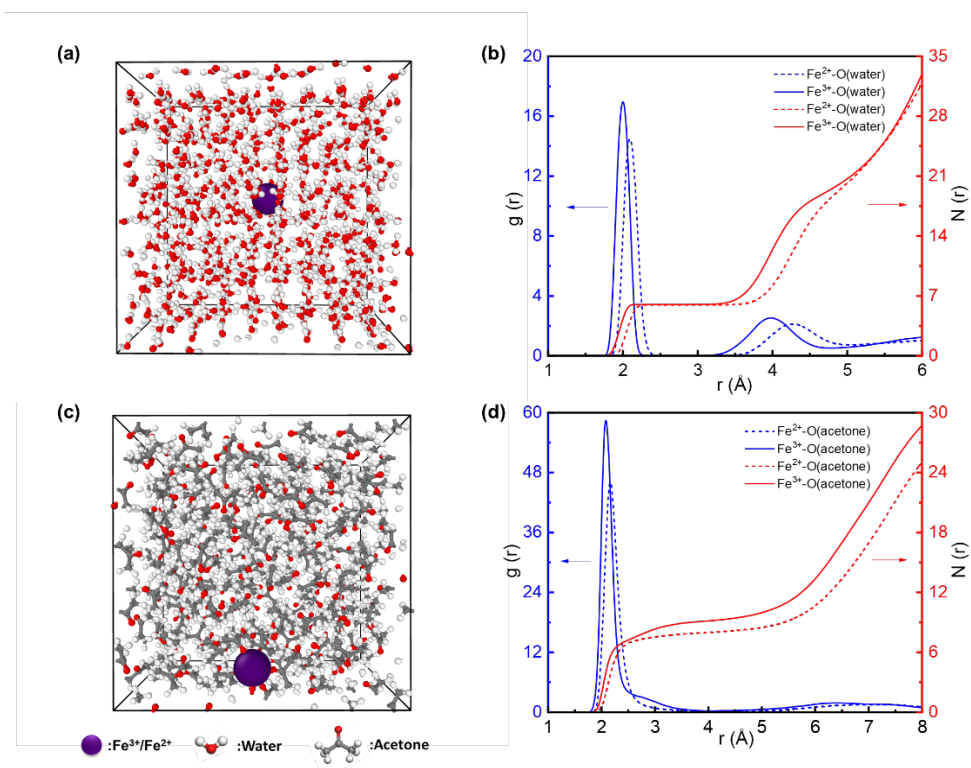

**Figure 2.** (a) Snapshot of water molecules and $Fe^{3+}/Fe^{2+}$ ion in the simulation box. (b) $g(r)$ and $N(r)$ of the O atoms around $Fe^{3+}$ and $Fe^{2+}$ ions in pure water. (c) Snapshot of acetone molecules and $Fe^{3+}/Fe^{2+}$ ion in the simulation box. (d) $g(r)$ and $N(r)$ of the O atoms around $Fe^{3+}$ and $Fe^{2+}$ ions in pure acetone.

In addition to the radial distributions, the orientation of solvent molecules is also an important factor for solvation structure that could affect the thermogalvanic temperature coefficient $\alpha$. The probability distribution of molecular dipole orientations in the first solvation shell is therefore calculated. To obtain the distribution of dipole orientations, we first record the trajectories of molecular dipoles and plot the orientations on a unit sphere, as shown in Figure 3a-b. The different coordination number is manifested in the dipole trajectories, with 6 clusters of dipole orientations on the unit sphere for $Fe^{3+}/Fe^{2+}$ in water and 8 clusters of dipole orientations for $Fe^{3+}/Fe^{2+}$ in acetone. Using these trajectories, we calculated the probability distribution of dipole orientations characterized by the angle $\theta$ between the dipole moment and the radial direction (Figure 3c-d). Clearly, the $\theta$ angle has a broader distribution in acetone solutions than in water. Intuitively, a larger change in the solvation structure before and after the redox reaction means a larger entropy change $\Delta S_{rxn}$ and a higher $\alpha$. Given that solvation entropy is a measure of solvation shell disorder, we introduce the variance of dipole fluctuation $\sigma_i^2[\theta]$ as a descriptor of solvation shell order of ion species $i$ ($Fe^{3+}$ or $Fe^{2+}$). The variance is simply calculated as:

$$\sigma_i^2[\theta] = \int (\theta - \bar{\theta})^2 p(\theta) \mathrm{d}\theta \tag{5}$$

with $\bar{\theta} = \int \theta p(\theta) \mathrm{d}\theta$ the mean orientation angle of dipoles and $p(\theta)$ the probability distribution function. The $\sigma_{Fe^{3+}}^2[\theta]$ and $\sigma_{Fe^{2+}}^2[\theta]$ in water solutions are calculated as 32.57 and 58.16, while these values in the acetone system are 32.79 and 70.76, respectively. A higher variance value means the solvent molecule has a higher magnitude of thermal fluctuations, indicating the solvation shell structure is less ordered. The difference between $\sigma_{Fe^{3+}}^2[\theta]$ and $\sigma_{Fe^{2+}}^2[\theta]$ is therefore a microscopic indicator of solvation shell order change due to the redox reaction, defined as:

$$\zeta_{rxn} = \sigma_{Fe^{2+}}^2[\theta] - \sigma_{Fe^{3+}}^2[\theta] \tag{6}$$

The relation between the solvation shell order change parameter $\zeta_{rxn}$ and the effective thermopower $\alpha$ is shown in Figure 3e. Clearly, higher $\zeta_{rxn}$ is consistent with the increased $\alpha$ in acetone solutions.

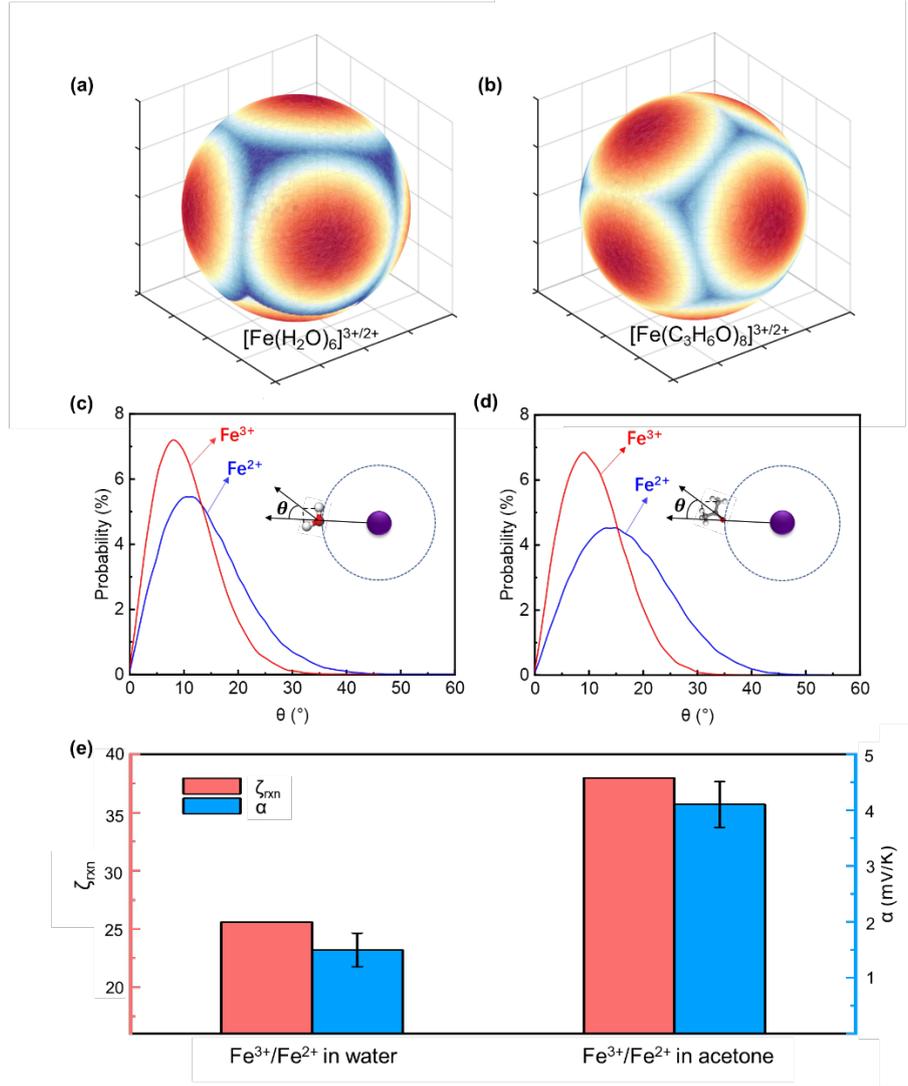

**Figure 3.** The trajectory of dipoles of (a) water and (b) acetone molecules in the first solvation shell of Fe ions. Each point on the sphere represents the dipole orientation of a solvent molecule at an MD snapshot. The color indicates the number density of sampled dipole orientations per solid angle, with red and blue colors representing high and low density, respectively. Angular probability distribution of solvent dipoles obtained from the MD simulations of the Fe ions in (c) pure water and (d) pure acetone. (e) $\zeta_{rxn}$ and the temperature coefficients $\alpha$ of $Fe^{3+}/Fe^{2+}$ in water and acetone solutions.

**Effect of acetone fraction on $\alpha$**

After studying the pure water and acetone solutions of $Fe^{3+}/Fe^{2+}$, the thermopower of electrolytes with mixed solvents is explored in this section, focusing on the effect of acetone fraction on the thermogalvanic temperature coefficient. Since redox ions typically have lower solubility in organic solutions, mixing the organic solvent with water could be favorable for the electrochemical performance of thermogalvanic cells[39].

We perform FEP computations and solvation structure analysis for mixed acetone-water solvent systems as shown in Figure 4a by varying the weight fraction $\phi_{wt}$ of acetone. As shown in Figure 4b, the effective thermopower $\alpha$ shows a two-stage dependence on $\phi_{wt}$. At low fractions $\phi_{wt} \leq 40\%$, the effective thermopower $\alpha$ is nearly independent of $\phi_{wt}$. However, when the acetone fraction $\phi_{wt}$ is higher than 60%, the effective thermopower $\alpha$ increases rapidly, with $\alpha$ of 2.57 mV/K, 3.14 mV/K, and 4.10 mV/K at $\phi_{wt}$ of 60%, 80% and 100%, respectively.

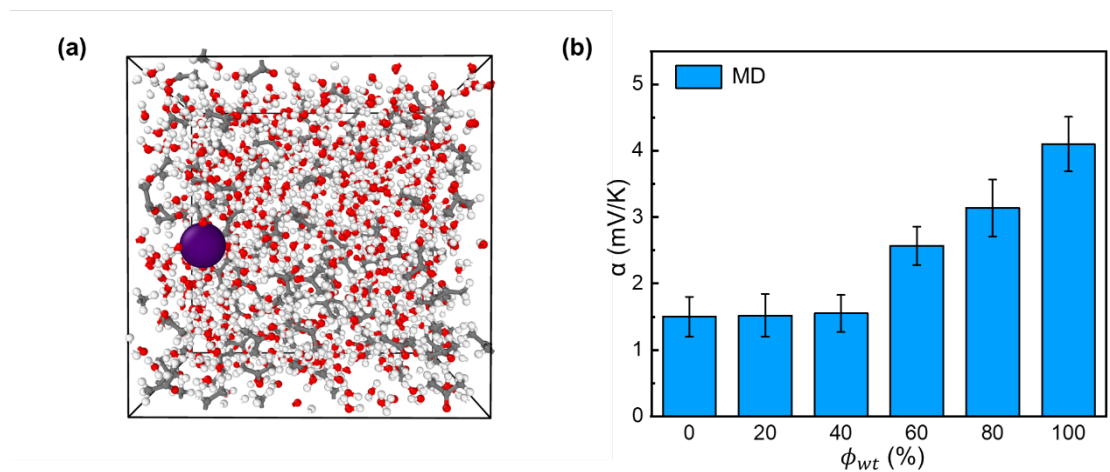

**Figure 4.** (a) Snapshot of $Fe^{3+}/Fe^{2+}$ ion in the mixed acetone-water solvent with $\phi_{wt}$ =40%. (b) Dependence of the temperature coefficients $\alpha$ on the acetone weight fraction $\phi_{wt}$.

To understand the increase of $\alpha$ with $\phi_{wt}$ at high acetone fractions, we studied the dependence of solvation shell structure on $\phi_{wt}$ for the mixed solvent systems. The radial distribution function and the coordination number of $Fe^{3+}/Fe^{2+}$ at 20% and 80% acetone fractions are shown in Figure 5a-d. RDF and CND at other acetone fractions are shown in Supplementary materials (Figure S1a-d). At low acetone fractions, the first peak in $g_{Fe-O}(r)$ is only contributed by water molecules, indicating that there are no acetone molecules inside the first solvation shell. In contrast, an oxygen peak from the acetone molecule emerges for both $Fe^{2+}$ ions (Figure 5b), when the acetone weight fraction $\phi_{wt} \geq 60\%$. Such phenomenon of acetone insertion into the first solvation shell is also manifested in CND $N(r)$. Within the first solvation radius $R_{1st}$, there are 6 water molecules in both $Fe^{2+}$ and $Fe^{3+}$ at the low acetone fraction. At the high acetone fractions, the water coordination number remains unchanged at 6 for the $Fe^{3+}$ ion, but the first solvation shell of the $Fe^{2+}$ ion contains 1 acetone molecule and only 5 water molecules (Figure 5e). The fact that the insertion of acetone molecules into the first solvation shell only happens at high acetone fractions $\phi_{wt} \geq 60\%$ can be explained by molecular polarity. Since water is a strong polar solvent, it is strongly confined by the electric field of the charge ions and more energetically favored to form the first solvation shell. At high acetone fractions, the lower-charged $Fe^{2+}$ ion has a weaker ionic field. It is thus more vulnerable to acetone insertion into the first solvation shell, and the probability for the $Fe^{2+}$ ion bounded with acetone is considerable at high acetone fractions.

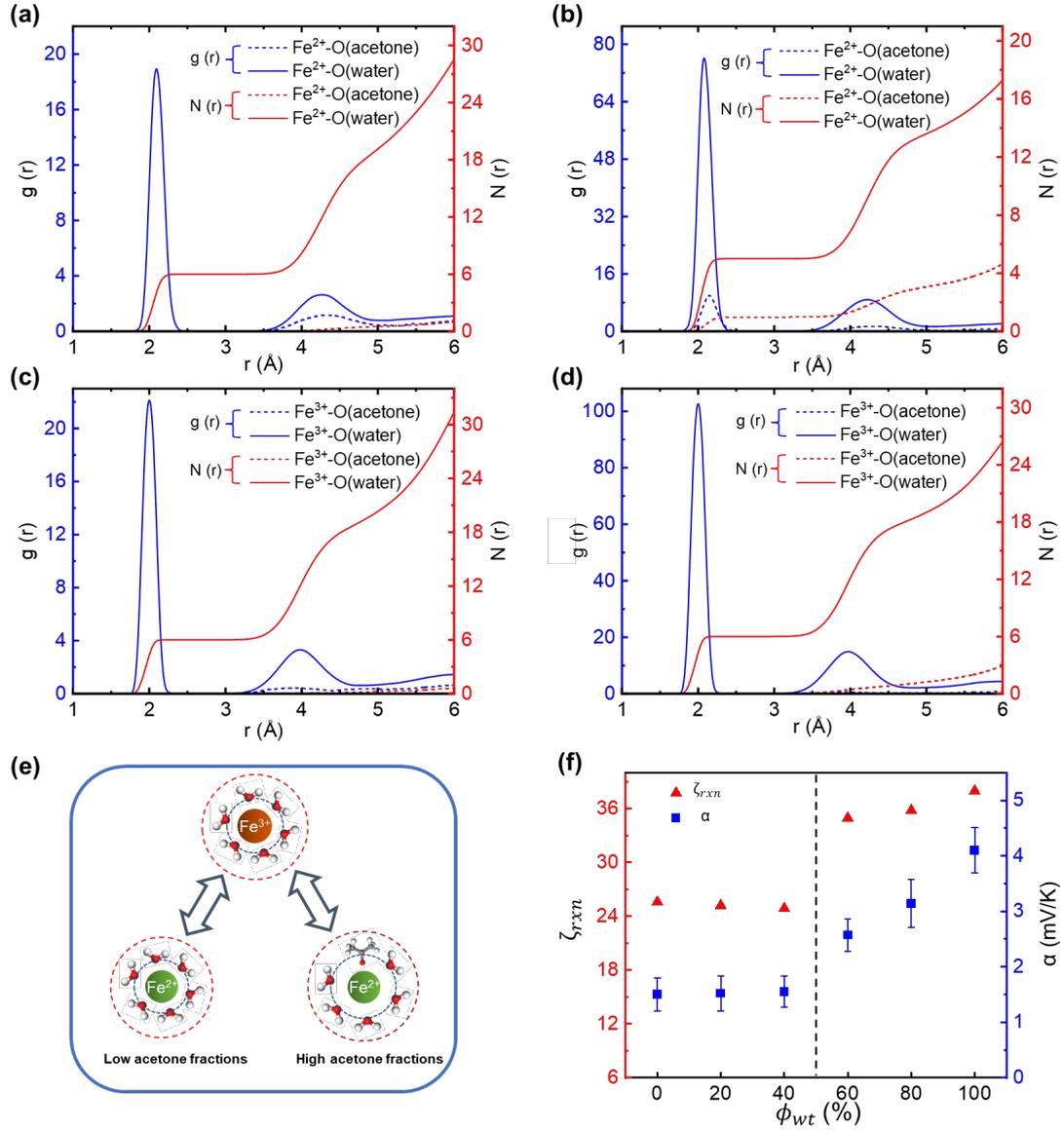

**Figure 5.** $g(r)$ and $N(r)$ of the O atoms from water and acetone around Fe ions in the mixed acetone-water solvent systems at different acetone weight fractions: $Fe^{2+}$ at (a) $\phi_{wt} = 20\%$ and (b) $\phi_{wt} = 80\%$, and $Fe^{3+}$ at (c) $\phi_{wt} = 20\%$ and (d) $\phi_{wt} = 80\%$. (e) The difference in first solvation shell structure after the redox reaction of $Fe^{3+}/Fe^{2+}$ between the low and the high acetone fraction systems. (f) Correlation between solvation order change $\zeta_{rxn}$ and the temperature coefficients $\alpha$ at different acetone weight fractions $\phi_{wt}$.

Since the acetone insertion into the first solvation shell also happens at $\phi_{wt} = 60\%$, it should be the key to explaining the rapid increase in $\alpha$ at high acetone fractions $\phi_{wt}$. To understand the effect of acetone insertion on solvation structure, we analyzed the correlation between $\alpha$ and solvation order change $\zeta_{rxn}$ at different acetone fractions $\phi_{wt}$. As shown in Figure 5f, $\zeta_{rxn}$ is nearly independent of $\phi_{wt}$, showing a similar acetone fraction dependence to $\alpha$. However, at high acetone fractions, $\zeta_{rxn}$ dramatically increases with $\phi_{wt}$ increasing from 60% to 100%, indicating that the acetone insertion effect in $Fe^{2+}$ solvation shells strongly increases the solvation structure order change of $Fe^{3+}/Fe^{2+}$ redox reaction, therefore resulting in a rapid boost of $\alpha$ with $\phi_{wt}$.

In summary, this work used atomistic simulations to predict the thermogalvanic temperature coefficients $\alpha$ of electrolytes, which is the key parameter of thermogalvanic batteries. The predicted $\alpha$ of $Fe^{3+}/Fe^{2+}$ shows nearly threefold enhancement from 1.5±0.3 mV/K in aqueous solutions to 4.1±0.4 mV/K in acetone solutions. By analyzing the solvation shell structure, we discovered that the difference in the probability distribution of solvent dipole orientations between $Fe^{3+}$ and $Fe^{2+}$ is much larger in acetone solutions than that in aqueous solutions, which is responsible for the enhanced $\Delta S_{rxn}$ and $\alpha$ in acetone solution. For acetone-water mixed solvents, we studied the effect of acetone fraction on temperature coefficient $\alpha$. Interestingly, only at high acetone fractions ($\phi_{wt} \geq 60\%$), thermopower $\alpha$ is increased by increasing acetone fractions $\phi_{wt}$, but $\alpha$ is almost independent of acetone weight fractions when $\phi_{wt} \leq 40\%$. This is because the acetone can insert into the first solvation shell of $Fe^{2+}$ at a high acetone fraction. We also found that the parameter $\zeta_{rxn}$ characterizing the solvation shell order change due to the redox reaction is a microscopic indicator of the magnitude of $\alpha$. This work provides microscopic insights into developing redox electrolytes with high thermogalvanic temperature coefficients.

## Methods

**Simulation structure and potential field**

In this work, MD simulations are performed using the Large Scale Atomic/Molecular Massively Parallel Simulator (LAMMPS) code[40]. All FEP and structure analyses are based on a cubic simulation box with a length of 31 Å containing a single $Fe^{3+}$ or $Fe^{2+}$ ion. The number of solvent molecules is added to the box to match the actual density with details listed in Supplementary materials (Table S1). The initial atomic coordinates were generated with Packing Optimization for Molecular Dynamics Simulation (Packmol) program[41]. Periodic boundary conditions (PBCs) were applied in all three directions for all simulations. A cutoff of 10 Å was used for both van der Waals interactions and long-range correction (particle-particle-particle-mesh, PPPM) of Coulombic interactions.

SPC/E rigid potential model[42] is selected to describe the interaction between water molecules, which has been widely applied for calculating thermodynamic properties such as the solvation free energy[42, 43]. For organic solvent acetone molecules, the Optimized Parameters for Liquid Simulations all-atom (OPLSS-AA) force field is used to parametrize the intermolecular forces[44], and the atomic charge is obtained from the 1.14*CM1A model[45]. Finally, Lennard-Jones (LJ) potential and Coulombic potential are used to describe the ion-solvent interactions. The LJ parameters of $Fe^{3+}/Fe^{2+}$ ions are obtained from the universal force field (UFF)[46], and geometric mixing rules to obtain the parameters for all other ion-solvent interaction pairs. Such parametrization strategy has been used in the previous literature[47], showing an excellent accuracy of reproducing the first solvation radius and RDF. Detailed force field parameters of ions and solvent molecules are shown in Supplementary materials (Table S2).

**FEP simulations**

We introduce the coupling parameter $\lambda$ into the ion-solvent interactions by adopting the soft-core potential function:

$$U(r) = 4\lambda\varepsilon\left(\frac{1}{\left[\alpha_{LJ}(1-\lambda)^2+\left(\frac{r}{\sigma}\right)^6\right]^2} - \frac{1}{\alpha_{LJ}(1-\lambda)^2+\left(\frac{r}{\sigma}\right)^6}\right) + \lambda\frac{q_i q_j}{[\alpha_C(1-\lambda)^2+r^2]^{\frac{1}{2}}} \quad (7),$$

where the first term represents the van der Waals interactions while the second term represents the Coulombic interactions; $q_i$ and $q_j$ represent the charge of ions and solvent atoms; $r$ denotes the distance between ions and solvent atoms; $\varepsilon$, $\sigma$ are the parameters of Lennard-Jones (LJ) potential; where the dimensionless coefficients $\alpha_{LJ} = 0.5$ and $\alpha_C = 10$ are typically used in the FEP method[48]. Before the solvation free energy calculation, the ion-solvent systems are first equilibrated using the isobaric-isothermal (NPT) ensemble for 2 ns with $\lambda = 0$. Then the coupling parameter $\lambda$ is gradually increased from 0 to 1 with 0.05 as the interval, and this FEP calculation process was performed for 4 ns. For each $\lambda$ value, the solvation free energy is calculated using Eq. (4) and averaged for 0.2 ns. We repeat the above process by decreasing $\lambda$ from 1 to 0, which corresponds to the de-solvation process, and the final solvation free energy is obtained by averaging the two results. At each temperature, 10 independent simulations with different initial velocity distributions are performed for a better ensemble average.

**Analysis of solvation structure**

The radial distribution of solvent molecules around the $Fe^{3+}/Fe^{2+}$ ions is described in terms of the radial distribution function (RDF) $g_{AB}(r) = \frac{\langle \rho_B(r) \rangle}{\langle \rho_B \rangle}$, where $\langle \rho_B(r) \rangle$ is the particle density of type B at distance $r$ around type A, and $\langle \rho_B \rangle$ is the average particle density of type B in the investigation region. This calculation was performed by the MD simulation in the NPT ensemble. First, the simulation cell is equilibrated for 5 ns using the Noosé-Hoover thermostat under the NPT ensemble. Subsequently, the simulation cell is run for another 10 ns for sampling RDF. In total, 20,000 snapshots separated by 500 fs are used to calculate the averaged RDF of solvent molecules around $Fe^{3+}/Fe^{2+}$ ions.

The orientation distribution of the solvent molecules is analyzed by computing the solvent dipole orientation in the first solvation shell. The dipole moment of the solvent molecules is calculated according to the formula $\boldsymbol{M}(t) = \sum_i z_i e \boldsymbol{r}_i(t)$, where $z_i$ and $\boldsymbol{r}$ are the valence and position of atom $i$, respectively. Similar to the sampling procedure used to compute RDF, molecular dipoles of solvent molecules are recorded every 500 fs, and we project $\boldsymbol{M}(t)$ to a unit sphere for visualizing the orientations as shown in Figure 4a-b. Probability distribution $p(\theta)$ is calculated by:

$$p(\theta) = \frac{1}{N_S N_{mol}} \frac{N_c(\theta, \theta + \Delta\theta)}{\Delta\theta} \quad (8)$$

where $N_S$ is the total number of sampled snapshots, $N_{mol}$ denotes the number of solvent molecules in the simulation cell, $N_c(\theta, \theta + \Delta\theta)$ is the number of dipoles within the interval $[\theta, \theta + \Delta\theta)$, and the bin size for computing probability distribution is $\Delta\theta = 1°$.

**Acknowledgments**

X. Q. acknowledges support from the National Natural Science Foundation of China (NSFC Grant No. 52276065). R.Y. acknowledges financial support from the National Key Research and Development Program of China (Grant No. 2022YFB3803900). The authors declare no conflict of interest. All simulations are performed using the High-Performance Computation Platform of Huazhong University of Science and Technology.

**Competing Interests**

The Authors declare no Competing Financial or Non-Financial Interests.

**Data Availability**

The data that supports this work can be found in the Manuscript and Supplemental Material. Additional data and computation scripts will be available from the corresponding author upon reasonable request.

**Author Contributions**

Y.C. and X.Q. performed MD simulations; Y.C. and T.H.L. performed structural analysis of the solvation shell; Y.C., Q.H, X.Q. and R.Y. wrote the manuscript; X.Q. and R.Y. conceptualized and supervised the research; all authors discussed the results and edited the manuscript.

# Supplementary materials:

# Effect of solvation shell structure on thermopower of liquid redox pairs

Yuchi Chen[1], Qiangqiang Huang[1], Te-Huan Liu[1], Xin Qian[1*] and Ronggui Yang[1,2*]

[1] School of Energy and Power Engineering, Huazhong University of Science and Technology, Wuhan 430074, China.

[2] State Key Laboratory of Coal Combustion, Huazhong University of Science and Technology, Wuhan 430074, China.

**Table S1.** Numbers of solvent molecules in the simulation systems with single ions ($Fe^{2+}/Fe^{3+}$), and the density of these systems.

| System | Acetone | Water | Density(MD)[a] | Density(Exp.)[b] |
|---|---|---|---|---|
| Pure water |  | 810 | 0.995 | 1.000 |
| Pure acetone | 250 |  | 0.779 | 0.788 |
| 20% acetone[c] | 60 | 750 | 0.956 | 0.958 |
| 40% acetone[c] | 110 | 550 | 0.920 | 0.915 |
| 60% acetone[c] | 160 | 340 | 0.881 | 0.873 |
| 80% acetone[c] | 205 | 160 | 0.842 | 0.830 |

[a]. The average density was calculated at 300 K temperature in this work. [b]. The experimental measurements of density at room temperature. [c]. It represents the mixed acetone-water solvent system at different weight fractions of acetone.

**Table S2.** Non-bonded parameters used in the MD simulations of ions and solvent molecules

| Ion/molecule | Atomic Site | q | ε (kcal/mol) | σ(nm) |
|---|---|---|---|---|
| $Fe^{2+}$ | Fe | +2.0 | 0.013 | 0.259 |
| $Fe^{3+}$ | Fe | +3.0 | 0.013 | 0.259 |
| $K^+$ | K | +1.0 | 0.035 | 0.340 |
| $Na^+$ | Na | +1.0 | 0.030 | 0.266 |
| $Cl^-$ | Cl | -1.0 | 0.227 | 0.352 |
| $I^-$ | I | -1.0 | 0.339 | 0.401 |
| Acetone | C(H)[a] | -0.3019 | 0.066 | 0.350 |
|  | C(O)[b] | 0.3627 | 0.070 | 0.355 |
|  | O | -0.4561 | 0.210 | 0.296 |
|  | H | 0.1162 | 0.030 | 0.250 |
| Water | O | -0.8476 | 0.1553 | 0.3166 |
|  | H | 0.4238 | 0.0000 | 0.0000 |

[a]. It represents the C atom attached to three H atoms in acetone. [b]. It represents the C atom attached to one O atom in acetone.

**Table S3.** The solvation entropy $\Delta S^{solv}$ ($10^{-3}$ eV/K) of redox ions ($Fe^{2+}$ and $Fe^{3+}$).

| System | $\Delta S^{solv}$ of $Fe^{2+}$ | $\Delta S^{solv}$ of $Fe^{3+}$ |
|---|---|---|
| 20% acetone | 3.74±0.24 | 5.26±0.56 |
| 40% acetone | 3.95±0.41 | 5.50±0.70 |
| 60% acetone | 2.70±0.57 | 5.27±0.86 |
| 80% acetone | 3.39±0.71 | 6.53±1.15 |
| Pure acetone | 4.35±0.72 | 8.49±1.13 |

Note: These results were used to calculate the temperature coefficient resulting in Figure 1d and Figure 4b.

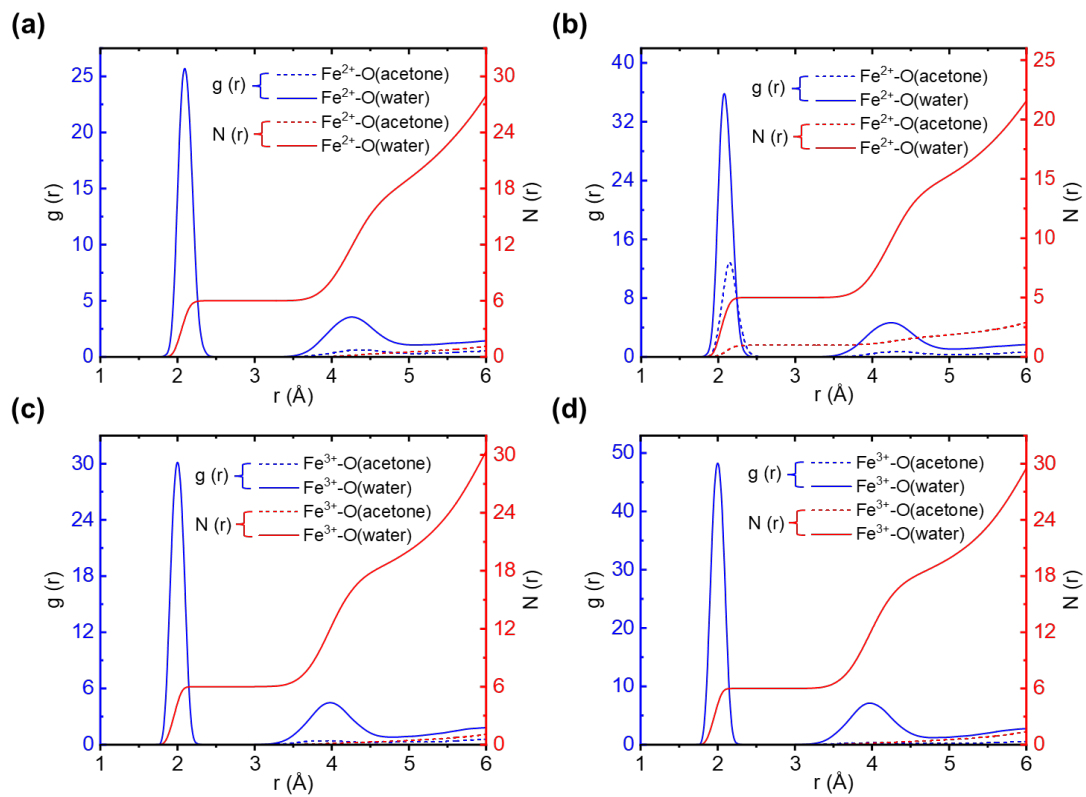

**Figure S1.** $g(r)$ and $N(r)$ of the O atom from water and acetone around Fe ions in the mixed acetone-water solvent systems at different acetone weight fractions: $Fe^{2+}$ at (a) $\phi_{wt} = 40\%$ and (b) $\phi_{wt} = 60\%$, and $Fe^{3+}$ at (c) $\phi_{wt} = 40\%$ and (d) $\phi_{wt} = 60\%$.

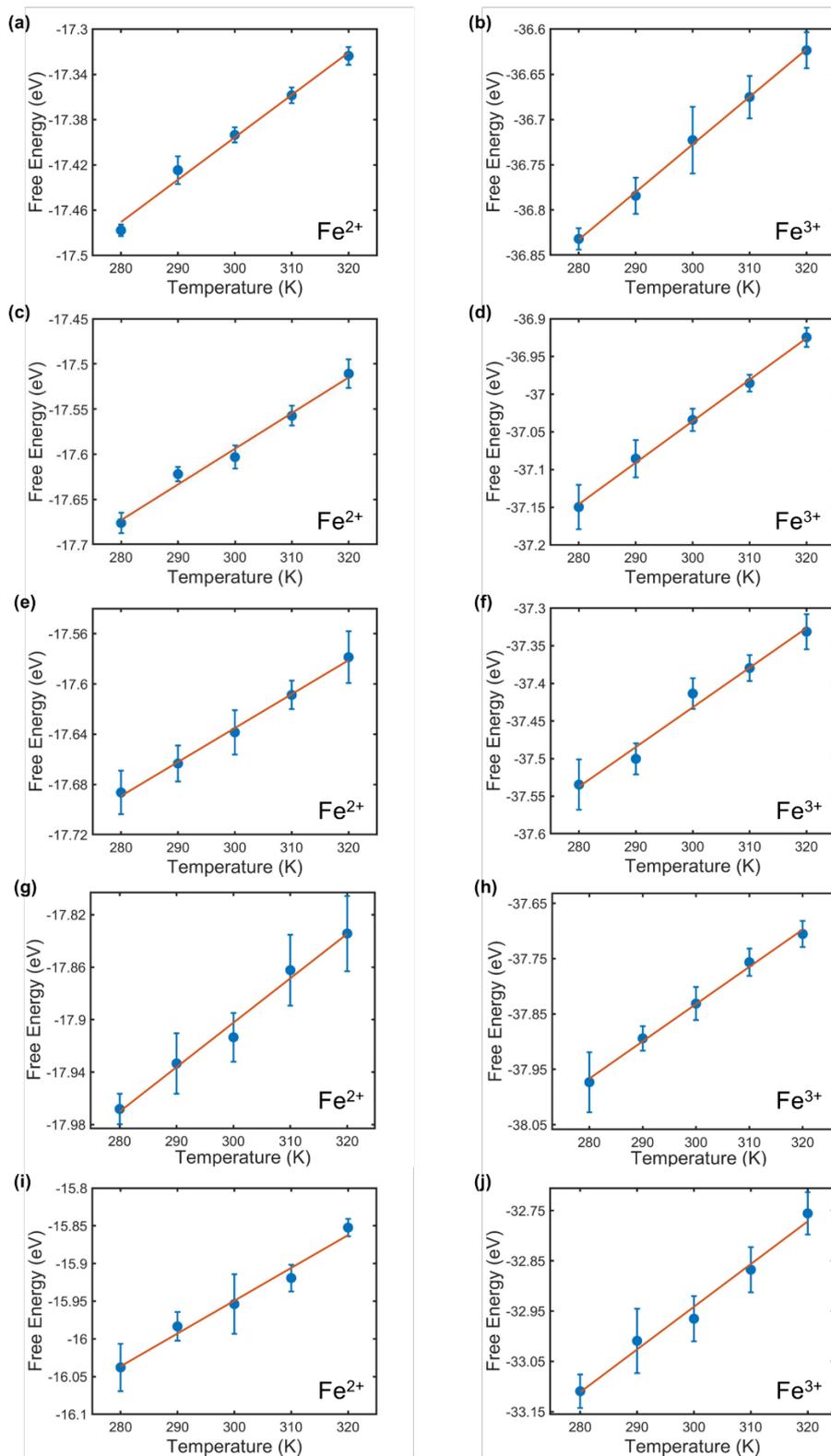

**Figure S2.** The solvation free energy $\Delta G^{solv}(T)$ of $Fe^{2+}$ and $Fe^{3+}$ in the mixed acetone-water solvent from 280 K to 320 K with acetone weight fractions of (a-b) $\phi_{wt} = 20\%$; (c-d) $\phi_{wt} = 40\%$; (e-f) $\phi_{wt} = 60\%$; (g-h) $\phi_{wt} = 80\%$; (i-j) $\phi_{wt} = 100\%$ (pure acetone). The solvation entropy $\Delta S^{solv}$ of redox ions ($Fe^{2+}$ and $Fe^{3+}$) obtained from the slope of the best linear fit is summarized in Table S3.